\documentclass[showpacs,preprintnumbers,amsmath,amssymb, 
eqsecnum,
twocolumn, tightenlines,
]{revtex4}

\usepackage{bm}

\usepackage{amsmath}

\usepackage{dcolumn}

   \usepackage[dvips]{graphicx}
\begin{document}

    \bibliographystyle{apsrev}
    
    \title {Tunneling into black hole, escape  from
      black hole, reflection from horizon and pair creation }

    \author{V.V.Flambaum}
    \email[Email:]{flambaum@phys.unsw.edu.au}
    
    \affiliation{School of Physics, University of New South Wales,
      Sydney 2052, Australia}
    
    \date{\today}

\begin{abstract}
  Within classical general relativity, a  particle cannot reach
the horizon of a black hole during a finite time, in the reference frame
of an external observer; a particle inside cannot escape from
a black hole; and the horizon does not produce any reflection.
We argue that these processes may possibly be allowed in the quantum world.
It is known that quantum mechanics allows pair creation at the horizon
(one particle inside, another particle outside) and
Hawking radiation. One can extend this idea to propose
other processes. Tunneling of an external
 particle inside black hole may be produced
by the creation of a pair at the horizon, followed by the annihilation
of  one created particle with the initial particle outside, with the
other created particle appearing inside. Escape of 
a particle from a black hole may result from the creation
of a pair, followed by the annihilation of one created particle with
the particle inside, with the
other created  particle appearing outside.
 The escape may allow the  transfer of  information to the outside.
Finally, the reflection of an external particle
from the horizon may be modelled by a combination of the two processes
 presented above.
The relationship between these ``pair creation-annihilation'' mechanisms and the
 ``horizon tunneling'' calculations \cite{kuchiev_1,kuchiev_2,kuchiev_3,
kuchiev_flambaum_04,kuchiev_flambaum_2} is discussed.
\end{abstract}
    
    \pacs{04.70.Dy, 04.20.Gz}

    \maketitle
    \section{Pair creation-annihilation model}    
    \label{pair}

As is known, a classical particle cannot  reach the horizon
of a black hole
during a finite time in the reference frame of an external observer.
This seems to forbid any obsevable effects which can be induced at the
horizon. For example, when the quantum scattering problem
was considered for black holes, the Matzner boundary condition
\cite{matzner} -- that there is no outgoing wave from the horizon
(complete absorption) -- was used. This makes the capture cross-section
finite (proportional to the horizon area) in the limit of zero energy
of the scattered particle \cite{starobinsky_73,starobinsky_churilov_73,
unruh_76} (see also \cite{sanchez_1977,sanchez_1997,frolov_novikov_98,
thorne_1994,chandrasekhar_93}).
However, recent works
 \cite{kuchiev_1,kuchiev_2,kuchiev_3} claim that the horizon
reflects low energy particles with probability $\exp(-\varepsilon/T)$
where $\varepsilon$ is the particle energy, and  $T$ is the Hawking temperature
of the black hole (for  black holes with zero charge and angular momentum
$T=\hbar c^3/8 \pi G M$). This makes the absorption cross-section zero for  
$\varepsilon=0$ \cite{kuchiev_flambaum_04}.
 This provokes a question: how can the ``unreachable''
horizon  produce so strong an effect?
 
   A naive answer to this question may refer to the uncertainty principle:
if the wavelength of an approaching particle is larger than the horizon
 radius, one cannot say where the particle is located relative to the
 horizon. One may even compare this case with a parabolic barrier,
$U=-kx^2$, for a particle with zero energy. A  classical particle
will climb this barrier only  during an infinite time.
Therefore, this is an impenetrable barrier for the incoming classical
 particle, similar to the black hole horizon. A quantum particle can either
 easily penetrate this barrier or be reflected from it.

 Another argument may be
based on the phenomenon of pair creation and Hawking radiation
near the horizon. A conventional explanation of the Hawking radiation
tells us that a pair is created near the horizon. The positive energy
particle escapes, and the  negative energy particle dives inside
the black hole, reducing its mass \cite{hartle_hawking_1976}.
 The Hawking radiation happens
on a finite time scale, in spite of the fact that a classical particle
cannot reach, or escape from, the horizon within a finite time.

  Using this pair creation picture we can build a qualitative model which
 demonstrates that a particle can disappear from our world (into black hole)
within a finite time, a particle can escape from a black hole
within a finite time, and there may be a reflected wave from the black
hole horizon. 

  Consider first an external particle approaching the horizon.
Then a pair is created by the strong gravitation field near the horizon.
If the horizon can produce a pair such that one particle is located at a
finite distance after finite time, it can also annihilate a pair at a
 finite distance, which is just the reverse process.
 Therefore, the initial
 external particle can annihilate with one of the created  particles
(``virtual antiparticle'') within a finite time. The other particle of
 the pair will be inside
the black hole. As a result, we have the initial particle
disappearing from our world within a finite time, a process which
is impossible for a classical particle. This means that the accretion
of matter into a black hole may also happen during finite time,
contrary to classical general relativity (although the accretion
process may take a longer time than the evaporation of the black hole).
 
   What factor determines the probability of this tunneling
process? The Hawking radiation is determined by the Boltzmann factor
 $\exp(-\varepsilon/T)$, 
where $\varepsilon$ is the particle energy, and $T$ is the Hawking temperature
of the black hole. A somewhat similar factor should appear in the tunneling
of an external particle, since this process  can be modelled by pair
creation and annihilation near the horizon (similar to the Hawking radiation
which is due to the pair creations).
 
  We can repeat this modelling for a particle
inside a black hole, providing this inner particle has a positive
energy and can tunnel into the outside world. The
 escape of a particle from a black hole may result from the creation
of a pair, followed by annihilation of one created particle with the particle
 inside, and the other created particle appearing outside. Again,
 the probability
should depend on  $\exp(-\varepsilon/T)$. This process creates
 a particle outside the  black hole, which is identical to the particle
that was inside. Therefore it reveals information about the
contents of the black hole. For example, we may conclude that
the black hole contains matter and no antimatter.
Thus, the escape process transfers information from the black
hole outside.

 Finally, we can model the reflection of an external particle from the
horizon. A pair is created, one particle  annihilates with the initial
particle, the other particle  plays the role of the reflected wave.
 If the horizon
cannot produce two particles outside,  we should consider the  creation
 of two pairs followed by the annihilation of two particles created
inside, i.e. the  combination of two processes described above.
  Again, the probability should depend on  $\exp(-\varepsilon/T)$.

     We have considered these pair creation-annihilation models only
to demonstrate the possibility of tunneling, escape,
and reflection in principle. In the works  \cite{kuchiev_1,kuchiev_2,kuchiev_3,
kuchiev_flambaum_04,kuchiev_flambaum_2} the quantitative results
for these effects were obtained
by solving the wave equation for a scalar particle in the gravitational
field of a black hole. However, in spite of the
 completely  different approach used in the present work, there is a certain
 similarity in the qualitative picture.
 The tunneling of an external particle
 into a  black hole, the escape of an inner particle from a black  hole,
 and the reflection
from the horizon 
  in  \cite{kuchiev_1,kuchiev_2,kuchiev_3,
kuchiev_flambaum_04,kuchiev_flambaum_2} are all determined
 by the factor $\exp(-\varepsilon/T)$.
For comparison, 
 let us present briefly
the derivation from Ref.\cite{kuchiev_flambaum_04}, where the capture
cross-section of a scalar particle by a black hole is calculated.
Note that the works \cite{kuchiev_1,kuchiev_2,kuchiev_3} contains
also other methods of derivation of the reflection from the black hole
horizon which give the same result for the reflection amplitude $R$.

\maketitle
\section{Scalar field near  the horizon of the
Schwarzschild black hole }    
\label{scalar}

    Consider the scalar field $\phi(x)$ in the vicinity of the
    Schwarzschild black hole with the metric
    \begin{equation}
      \label{schw}
      ds^2 = -\left(1-\frac{1}{r}\right)dt^2 + \frac{dr^2}{1-1/r}
    +r^2 d\Omega^2~,
    \end{equation}
    where $d \Omega^2 = d \theta ^2 + \sin^2 \theta d\varphi^2$. We use
    relativistic units $\hbar=c=1$  and 
    the condition $2GM=1$, where  $G$ is the gravitational constant and
     $M$ is the mass
    of the black hole. The Schwarzschild radius in these units
    is $r_g \equiv 2GM = 1$.  The Klein-Gordon equation
    $-\partial_\mu ( \sqrt{-g} \,g^{\mu\nu} \partial_\nu \phi) =
    \sqrt{-g} \, m^2\,\phi$ for the field $\phi(x)$ in the
    Schwarzschild metric allows the separation of variables $\phi(x) =
    \exp(-i \varepsilon t)Y_{lm} (\theta,\varphi) \phi_{l}(r)$, where
    $\varepsilon,l,m$ are the energy, the momentum and its projection.
    The  radial function $\phi_{l}(r)$  satisfies the equation
\begin{eqnarray}
        \label{phi''}
&&\phi''_{l} 
+ \left(\frac{1}{r}+\frac{1}{r-1} \right) \phi'_l
\\ \nonumber && +
\left(   p^2 +  \frac{\varepsilon^2+
    p^2}{r-1}+\frac{\varepsilon^2}{(r-1)^2}
-\frac{ l(l+1) }{r(r-1)} \right) \phi_{l} = 0~.
      \end{eqnarray}
      Here $p$ is the momentum at infinity. 

 The solution for an incoming
   wave in the outside region in the close vicinity of the horizon is
   $\phi_\mathrm{in}(r) = \exp[-i\varepsilon \ln(r-1)\,]$ for $r
   \rightarrow 1,~r>1$.  We continue it into the interior region $r<1$,
   using an analytical continuation over the variable $r$ into the
   lower semiplane of the complex plane of $r$
($\ln(r-1)=\ln(1-r)-i\pi$).  This procedure allows
   one to find the incoming wave in the interior region in the
   vicinity of the horizon, $r\rightarrow 1,~r<1$,
   \begin{equation}
     \label{interior1}
   \phi_\mathrm{in}(r) = |{\cal R}|^{1/2}   \exp[ -i\varepsilon
   \ln(1-r)\,]~.
     \end{equation}
     It is suppressed compared to the outside region by a factor
$|{\cal R}|^{1/2}$ where 
     $|{\cal R}| = \exp(-2\pi \varepsilon)= \exp[-\varepsilon/(2T)$
   and  $T$ is the Hawking temperature. We may interprete  $|{\cal R}|$
as the probability of a particle tunneling into a black hole.

Continue now the
     incoming wave further into the interior region, $r<1$, using the
     differential equation (\ref{phi''}). In the vicinity of the
     origin, $r\rightarrow 0$, the solution
     can be presented as
   \begin{equation}
     \label{interior3}
   \phi_\mathrm{in}(r) = u \ln r + v~,
     \end{equation}
     where $u,v$ are constants. We assume that the total
     wave function should satisfy the conventional regular condition
     $\phi(r) \rightarrow const$ at the origin. Since the incoming
     wave exhibits a singular ($\propto \ln r$) behavior,
     there should exist an outgoing wave that compensates the
     singularity in the vicinity of the origin. Thus, the regular solution
is the standing wave.

 Repeating now the
     arguments in the reverse order, we take this outgoing wave and
     continue it towards the horizon using the differential equation;
     then continue it over the horizon using the analytical
     continuation via the lower semiplane of the complex $r$-plane.
     As a result, the outgoing wave appears in the outside region
     as described by the second term in equation below:
\begin{equation}\label{reflection}
\phi_l(r) \rightarrow \exp[-i \varepsilon \ln(r-1)\,]
+ {\cal R}\exp[\,i \varepsilon \ln(r-1)\,]~.
    \end{equation}
Thus, the probabilty of the reflection is
 $P=|{\cal R}|^2 =\exp[-\varepsilon/(T)$. 
 
  Here we see a certain similarity of the results with the pair
 creation-annihilation
model considered in the first section. In both approaches the probabilities
 depend exponentially
   on $\varepsilon/T$.

\section{Conclusion}

The aim of this paper is to present some qualitative arguments 
that some of the famous conclusions  of  classical general relativity
 related to black holes may be, in principle,  questioned by 
 quantum mechanics. 
We considered a model where the tunneling, escape and reflection processes
are mediated by  the creation and annihilation of pairs at the black hole
horizon.
This model hints that a particle can disappear
from our world into a black hole within a  finite time.
Indeed,  it is known from the derivation of the Hawking radiation
  \cite{hartle_hawking_1976}, 
that the creation of a pair near the horizon (one particle outside a black hole, the other particle
inside) happens on a finite time scale. Therefore,  it would be natural
if a quantum particle can cross the horizon within a finite time.
Thus, the accretion
of matter into a black hole may happen during a finite time,
contrary to the classical general relativity.

 Similarly, a particle captured
inside a black hole can tunnel into the external world. This escape may
 transfer
some information outside. For example, we may find that the black hole
contains matter and no antimatter. The probability of the tunneling, escape,
and reflection processes should depend on the Boltzmann factor
 $\exp(-\varepsilon/T)$,
where $\varepsilon$ is the particle energy, and $T$ is the Hawking temperature.

   The author is grateful to M. Kuchiev, Y. Augarten and J. Berengut
for useful comments. This work was supported by the Australian Research Council.


\begin{thebibliography}{99}

  \bibitem{kuchiev_1} 
    
    M. Yu. Kuchiev, gr-qc/0310008.


  \bibitem{kuchiev_2}

  M. Yu. Kuchiev, Europhys. Lett. {\bf 65}, 445 (2004).


  \bibitem{kuchiev_3}
    
    M. Yu. Kuchiev, Phys. Rev. D {\bf 69}, 124031 (2004).


  \bibitem{kuchiev_flambaum_04}
    
    M. Yu. Kuchiev and V. V. Flambaum, gr-qc/0312065;
 Phys. Rev. D, Scheduled  15 Aug (2004).

\bibitem{kuchiev_flambaum_2}
    
    M. Yu. Kuchiev and V. V. Flambaum, gr-qc/0407077.


    \bibitem{matzner}
    J. A. H. Futterman, F. A.  Handler, and R. A.  Matzner,  {\it
      Scattering from black holes} (1988) Cambridge; New York:
    Cambridge University Press.


  \bibitem{starobinsky_73} 
    
    A. A. Starobinsky, Zh. Eksp. Teor. Fiz.  {\bf 64}, 48 (1973)
    [Sov.Phys.JETP {\bf 37}, 28 (1973)].

  \bibitem{starobinsky_churilov_73} 
    
    A. A. Starobinsky and C. M. Churilov, Zh. Eksp. Teor. Fiz.  {\bf 65},
    3 (1973) [Sov.Phys.JETP {\bf 38}, 1 (1974)].

  \bibitem{unruh_76}
    
    W. G. Unruh, Phys. Rev. D {\bf 14}, 3251 (1976); thesis Princeton
    Univ., 1971 (unpublished) (available University Microfilms, Ann
    Arbor, Mich.).

  \bibitem{sanchez_1977}
    
    N. Sanchez, Phys. Rev. D {\bf 18}, 1030 (1977).

  \bibitem{sanchez_1997}

    N. Sanchez, hep-th/9711068.


   \bibitem{frolov_novikov_98}
      
     V. P. Frolov and I. D. Novikov, {\it Black hole physics: basic
       concepts and new developments } (1998) Dordrecht; Boston: Kluwer.


  \bibitem{thorne_1994}
        
    K. S. Thorne, {\it Black hole and time warps} (1994) New York:
    Norton.
    
      
  \bibitem{chandrasekhar_93}
      
    S. Chandrasekhar, {\it The Mathematical Theory of Black Holes}
    (1993) New York: Oxford University Press.


  \bibitem{hartle_hawking_1976}
      
    J. B. Hartle and S. W. Hawking, Phys. Rev. D{\bf 13}, 2188 (1976).
S.W. Hawking, Phys. Rev. D{\bf 14}, 2460 (1976).

  \end{thebibliography}
\end{document}